\documentclass[prl,twocolumn,showpacs,preprintnumbers,amsmath,amssymb,times]{revtex4}

\usepackage{graphicx}% Include figure files
\usepackage{dcolumn}% Align table columns on decimal point
\usepackage{bm}% bold math
\usepackage{psfrag}
\usepackage{epsfig}
\usepackage{amsmath}
\usepackage{amssymb}
\usepackage{color}
\usepackage{fancyheadings}
\usepackage{mathbbol}

\newcommand{\bra}{\langle}

\newcommand{\ket}{\rangle}

\begin{document}

%\preprint{APS/123-QED}

\title{Strong photon non-linearities and photonic Mott insulators}

\author{Michael J. Hartmann}
\email{m.hartmann@imperial.ac.uk}
\author{Martin B. Plenio}
\affiliation{Institute for Mathematical Sciences, Imperial College London,
53 Exhibition Road, SW7 2PG}
\affiliation{QOLS, Blackett Lab., Imperial College London, Prince Consort Rd.,
SW7 2BW, United Kingdom}

\date{\today}

\begin{abstract}
We show, that photon non-linearities in electromagnetically induced transparency
can be at least one order of magnitude larger than predicted in all previous approaches.
As an application we demonstrate that, in this regime they give rise to very strong
photon - photon interactions which are strong enough to make an experimental realization of a
{\em photonic} Mott insulator state feasible in arrays of coupled ultra high-Q micro-cavities.
\end{abstract}

\pacs{03.67.Mn, 42.50.Dv, 73.43.Nq, 03.67.-a}% PACS, the Physics and Astronomy
%Classification Scheme.
\maketitle

% ---------------------------------------------------------------------------
%
\paragraph{Introduction:} 

Various quantum mechanical effects have been observed with photons.
Examples include quadrature squeezing and entangled photon states \cite{WallsMilburn}.
All these effects require strong non-linear interactions between photons for their observation.
Unfortunately, photons tend to interact only weakly and it is thus of considerable interest to develop schemes achieving larger and larger photon non-linearities, as these
would make further quantum effects accessible in experiments.

A scheme for the generation of a large photon non-linearity has been proposed by Imamo\u{g}lu and co-workers \cite{ISWD97}
for atoms with a level structure considered in electromagnetically induced transparency (EIT) \cite{FIM05},
which interact with the resonant light mode of a cavity and in \cite{GWG98} it was clarified
when this is strictly a photon-photon interaction.
Non-linearities obtained in this way can be strong enough to employ them as single photon
turnstile devices \cite{BBM+05}. The photon non-linearity considered in \cite{ISWD97} can be studied
in a more general regime as a non-linear interaction for dark state polaritons \cite{HBP06}.
This non-linearity and other approaches \cite{ASB06} have been employed to show that effective
many body physics should be observable in arrays of coupled cavities \cite{HBP06}.

In this letter, we show, that the photon non-linearity proposed in \cite{ISWD97} exists in a much more general parameter regime than previously considered \cite{GWG98}. In that way, we can relax a restricting assumption used in
\cite{ISWD97,GWG98} and \cite{HBP06} and are thus able to increase achievable photon-photon interactions by
at least one order of magnitude.
As the non-linear interactions presented here are significantly stronger than in previous schemes
they open up possibilities for the observation of phenomena which were
previously not accessible in experiments.
If generated in an array of coupled ultra high-Q micro-cavities \cite{AKS+03}, they 
can become large enough to make the experimental observation
of a Mott insulator state for photons rather than polaritons feasible.

The parameter regime we consider in this paper is crucially different to the dark state polariton
regime employed in \cite{HBP06} and the regime addressed in \cite{ASB06}.
Indeed, the non-linearity we derive here is an interaction between photons and not polaritons (joint atomic - photonic excitations) as considered in \cite{ISWD97,GWG98,HBP06,ASB06}.
 
In the photonic Mott state we consider, exactly one photon exists in each cavity,
provided the whole structure contains on average one photon per cavity \cite{GME+02}. Moreover, the photons are localized in the cavity they are in and are not able to hop between different cavities.
In such a situation photons behave as strongly correlated particles that 
are each "frozen" to their lattice site, a system that
would correspond to a crystal formed by light. The non-linearity derived here, when applied to coupled cavities,
could thus provide the first realistic possibility to observe light 
in this exotic quantum state. 

As this photonic Mott insulator is characterized by
the number of photons in a single cavity being, $\bra a_l^{\dagger} a_l \ket \approx 1$ for cavity $l$, and its fluctuations being zero,
$\bra (a_l^{\dagger} a_l)^2 \ket - \bra a_l^{\dagger} a_l \ket^2 \approx 0$,
it is fundamentally different to polaritonic Mott insulators
as predicted in \cite{HBP06} and \cite{ASB06}, where the number of polaritons at one site is exactly unity but the photon number fluctuates ($\bra a_l^{\dagger} a_l \ket \approx \frac{1}{2}$ and
$\bra (a_l^{\dagger} a_l)^2 \ket - \bra a_l^{\dagger} a_l \ket^2 \approx \frac{1}{2}$ for no detuning).

This letter provides two main results: We generalize the photon-photon interaction
\cite{ISWD97} showing that much stronger interactions can be obtained and, as an example for the
usefulness of this improvement, we demonstrate that with this new tool,
the experimental observation of a Mott insulator made of photons/light is within reach with
present day high-Q micro-cavities.  

\begin{figure}
\psfrag{g13}{\hspace{-0.04cm}$g_{13}$}
\psfrag{g24}{\hspace{0.5cm}$g_{24}$}
\psfrag{o}{\hspace{0.04cm}$\Omega$}
\psfrag{d}{\hspace{-0.14cm}$\Delta$}
\psfrag{d2}{\hspace{-0.cm}$\delta$}
\psfrag{e}{$\varepsilon$}
\psfrag{w1}{\hspace{-0.0cm}$\omega$}
\psfrag{w2}{\hspace{-0.0cm}$\omega$}
\psfrag{1}{$1$}
\psfrag{2}{\raisebox{-0.06cm}{$2$}}
\psfrag{3}{$3$}
\psfrag{4}{$4$}
\includegraphics[width=6cm]{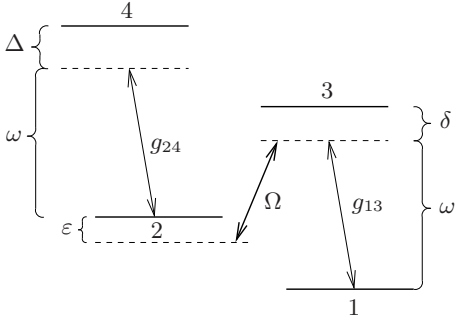}
\caption{\label{level} The level structure and the possible transitions of one atom, $\omega$ is the frequency of the cavity mode,
$\Omega$ is the Rabi frequency of the driving by the laser, $g_{13}$ and $g_{24}$
are the parameters of the respective dipole couplings and $\delta$, $\Delta$ and $\varepsilon$
are detunings.}
\end{figure}

\paragraph{Photon-photon interaction:} We begin by considering one cavity that interacts
with 4-level atoms which are driven by an external laser
in the same manner as in EIT, see fig. \ref{level}:
The transitions between levels 2 and 3 are coupled to the laser
and the transitions between levels 2-4 and 1-3 couple via dipole moments to the cavity mode.
The existence of a strong non-linearity in this atom cavity system was shown in \cite{ISWD97,GWG98} and a similar nonlinearity has recently been observed experimentally \cite{BBM+05}. 

In a rotating frame with respect to $H_0 = \omega ( a^{\dagger} a + \frac{1}{2}) + \sum_{j=1}^N ( \omega \sigma_{22}^j + \omega \sigma_{33}^j + 2 \omega \sigma_{44}^j )$, where we set $\hbar = 1$ throughout the paper, the Hamiltonian of the atoms in the cavity reads,
$H_I = \sum_{j=1}^N ( \varepsilon \sigma_{22}^j + \delta \sigma_{33}^j + (\Delta + \varepsilon)
\sigma_{44}^j )
+ \sum_{j=1}^N ( \Omega \, \sigma_{23}^j \, + \,
g_{13} \, \sigma_{13}^j \, a^{\dagger} \, + \,
g_{24} \, \sigma_{24}^j \, a^{\dagger} \, + \, \text{h.c.} ) \,$,
where $\sigma_{kl}^j = | k_j \ket \bra l_j |$ projects level $l$ of atom $j$ to level $k$ of the same atom,
$\omega$ is the frequency of the cavity mode,
$\Omega$ is the Rabi frequency of the driving laser and $g_{13}$ and $g_{24}$
are the couplings of the cavity mode to the respective atomic
transitions which are all assumed to be real. $\delta$, $\Delta$ and $\varepsilon$
are detunings of atomic transitions with respect to the cavity and laser fields, see fig. \ref{level}.
All atoms interact in the same way with the cavity mode and hence the only
relevant states are symmetric Dicke type dressed states\footnote{If the atoms were distributed on fixed positions in space, the dressed states are no longer symmetric but the approach still works exactly the same.}.
 
When the system is operated in the strong coupling regime, atoms and photons do no longer behave as
separate entities and polaritons, the excitations of dressed states, become a useful and intuitive
picture for its description. 
For the derivation, we thus make use of the polariton picture considered in \cite{HBP06} but will later consider the photonic limit. Hence while the technique is similar the physical conditions are different
to those considered in \cite{HBP06}. We define the polariton creation operators \cite{ISWD97},
$p_0^{\dagger}  = B^{-1} \, \left(g S_{12}^{\dagger} - \Omega a^{\dagger} \right)$ and
$p_{\pm}^{\dagger} = \sqrt{\frac{2}{A (A \pm \delta)}} \, \left(\Omega S_{12}^{\dagger} + g a^{\dagger} \pm
\frac{A \pm \delta}{2} S_{13}^{\dagger} \right)$
where $g = \sqrt{N} g_{13}$, $B = \sqrt{g^2 + \Omega^2}$, $A = \sqrt{4 B^2 + \delta^2}$,
$S_{12}^{\dagger} = \frac{1}{\sqrt{N}} \sum_{j=1}^N \sigma_{21}^j$
and $S_{13}^{\dagger} = \frac{1}{\sqrt{N}} \sum_{j=1}^N \sigma_{31}^j$.
The operator $p_0^{\dagger}$ excites the dark state in which level 3 is not occupied because both ways to
excite it interfere destructively \cite{FIM05}. Note in particular that in the limit $\Omega \gg g$, the dark state polaritons $p_0^{\dagger}$ become photons.

Neglecting the coupling to level 4, $g_{24} = 0$, and two photon detuning, $\varepsilon = 0$, the Hamiltonian $H_I$ in the relevant subspace spanned by Dicke type dressed states is given by
$\left[H_I\right]_{g_{24} = 0, \varepsilon = 0} = 
\mu_0 \, p_0^{\dagger} p_0 + \mu_+ \, p_+^{\dagger} p_+ + \mu_- \, p_-^{\dagger} p_- \, $,
with the polariton frequencies $\mu_0 = 0$ and $\mu_{\pm} = (\delta \pm A)/2$
Hence $p_0^{\dagger}, p_+^{\dagger}$ and $p_-^{\dagger}$ decouple for
$\varepsilon = g_{24} = 0$.

In a regime where $\Delta, g_{24} \ll \Omega$, only the dark state polaritons $p_0^{\dagger}$ couple to the atomic level 4 as for the other species their separation in energy from level 4 is much larger than their coupling to it. Hence the coupling
$a^{\dagger} \, \sum_{j=1}^N \sigma_{24}^j + \text{h.c.}$ can be approximated to read
$- g_{24} \, g \, \Omega \, B^{-2} \, S_{14}^{\dagger} \, p_0^2 + \text{h.c.}$ \cite{HBP06}.
In the limit $g \ll \Omega$ we can furthermore expand the coupling in powers of $g$ to obtain
$- g_{24} \, g \, \Omega^{-1} \, S_{14}^{\dagger} \, a^2 + \text{h.c.}$.
Provided that $g_{24} \, g \, \ll \, | \Delta \, \Omega |$, this coupling can be treated
perturbatively, yielding a 2nd order energy shift of
$n_{\text{ph}} \, (n_{\text{ph}} - 1) \, g_{24}^2 \, g^2 \, \Omega^{-2} \, \Delta^{-1}$,
where $n_{\text{ph}}$ is the number of photons in the cavity.
Note in particular that the assumption $g_{24} \ll \Delta$, which has been used in earlier
approaches \cite{ISWD97,HBP06}, is not made here. This assumption is not necessary since the atomic level 2
has very small occupation in the limit $\Omega \gg g$ and the perturbative treatment is justified even if
$g_{24} > \Delta$. This gives rise to photon non-linearities that can be orders of
magnitude larger than in previous approaches \cite{ISWD97,HBP06}.
The photon non-linearity can be expressed as the effective Hamiltonian
$- g_{24}^2 \, g^2 \, \Delta^{-1} \, \Omega^{-2} \, a^{\dagger} \, a^{\dagger} \, a \, a$ .
Note in particular that this photon-photon interaction is repulsive for $\Delta < 0$ and
attractive for $\Delta > 0$, giving easy experimental access to both regimes.
An analogous derivation shows that the two photon detuning $\varepsilon$ (see fig. \ref{level})
leads to a shift of the photon frequency given by the effective Hamiltonian
$\varepsilon \, g^2 \, \Omega^{-2} \, a^{\dagger} \, a$, which acts as a chemical potential.

The more the dark state polaritons become truly photons ($g \ll \Omega$), the larger is the
gain in non-linearity compared to previously considered scenarios \cite{ISWD97,HBP06}.
If we assume $\Omega = 100 g$ and $| \Delta | = g_{24} / 10$, the requirements $g \ll \Omega$
and $g_{24} \, g \, / \, | \Delta \, \Omega | = 1 / 10 \ll 1$ hold.
If one followed previous derivations \cite{ISWD97,HBP06}, one would need $| \Delta | \gg g_{24}$
and therefore choose $| \Delta | = 10 g_{24}$ or larger. Since the non-linearity %(\ref{nonlinearity})
is proportional to $1 / \Delta$, it is in our approach a factor $100$ larger than in previous ones.

An effective decay rate $\Gamma$ for the photons can be obtained following a similar derivation as in
\cite{HBP06} and only keeping leading terms in $g/\Omega$:
$\Gamma = \Gamma_C \, + \, \Theta(n_{\text{ph}}-2) \, g_{24}^2 \, g^2 \, \Delta^{-2} \, \Omega^{-2} \, \Gamma_4$ ,
where $\Theta(n_{\text{ph}}-2)$ is the Heaviside step function, $\Gamma_C$ the bare cavity decay rate and
$\Gamma_4$ the rate for spontaneous emission from level 4, which only
occurs if 2 or more photons are present in a cavity and is further suppressed since
$g_{24} g \ll |\Delta \Omega|$. Cavity decay is thus the main source of photon loss \footnote{To avoid super radiance the atomic levels are chosen such that the 4-1 transition is dipole forbidden.}.
Hence, although the non-linearity achieved here is much larger than in \cite{ISWD97,HBP06},
it does not come at the expense of stronger decay rates.

\paragraph{Photonic Mott insulator:} We now consider a periodic array of cavities. The resonant modes of neighboring cavities have a nonvanishing overlap which gives rise to photon hopping between them \cite{HRP06}. Including the on-site potential, the complete Hamiltonian for
$\varepsilon = 0$ takes on the form of a Bose-Hubbard (BH) Hamiltonian.
\begin{equation} \label{bosehubbard}
H_{\text{eff}} = U \sum_{\vec{R}} \left(a_{\vec{R}}^{\dagger}\right)^2 \left(a_{\vec{R}}\right)^2 +
J \sum_{\bra \vec{R}, \vec{R}' \ket} \left(
a_{\vec{R}}^{\dagger} \, a_{\vec{R}'} + \text{h.c.} \right)
\end{equation}
Here $\sum_{\bra \vec{R}, \vec{R}' \ket}$ is the sum of all pairs of cavities which are nearest neighbors of each other,
$J$ is given by an overlap integral over neighboring cavity modes
and can be obtained numerically for specific models, $a_{\vec{R}}^{\dagger}$ creates a photon in the cavity at site $\vec{R}$ and $U = g_{24}^2 \, g^2 \, \Delta^{-1} \, \Omega^{-2}$
is the on-site potential.
For successfully observing a Mott insulator state, the repulsion term $U$
needs to be much larger than the damping $\Gamma$. This can for example be the case for
toroidal micro-cavities \cite{AKS+03}, for which the achievable parameters are predicted to be
$g_{24} \sim 2.5 \times 10^{9} \text{s}^{-1}$,
$\Gamma_4 \sim 1.6 \times 10^{7} \text{s}^{-1}$ and
$\Gamma_C \sim 0.4 \times 10^{5} \text{s}^{-1}$. Taking
$g_{24} g / |\Delta \Omega| = 0.1$ and $g / \Omega  = 0.1$, we obtain $U / \Gamma \sim 625$.

To provide evidence for the accuracy of our approach and to show that the repulsion $U$ can indeed be strong enough to observe a Mott insulator state for photons, we present a numerical simulation of the
dynamics of 3 photons in 3 coupled toroidal micro-cavities that can tunnel between neighboring cavities according to $J (a_1^{\dagger} \, a_2, + \, \text{h.c.})$
($a_1^{\dagger}$ creates a photon in cavity 1 etc.) and interact with atoms in each cavity. We take parameters, which are predicted to be achievable in \cite{AKS+03}, and assume
$\delta = 1.0 \times 10^{11} \text{s}^{-1}$,
$\Delta = - 1.25 \times 10^{9} \text{s}^{-1}$,
$g_{13} = 2.5 \times 10^{9} \text{s}^{-1}$,
$N = 1000$, $\varepsilon = 0$,
$J = 1.2 \times 10^{6} \text{s}^{-1}$ and
$\Omega = 20 \sqrt{N} g_{13} = 1.58 \times 10^{12} \text{s}^{-1}$, which yields
$U = 1.24 \times 10^{7} \text{s}^{-1}$.
The initial state is 1 photon in each cavity, i.e. $| \psi_0 \ket = | 1, 1, 1 \ket$.
Cavity decay and spontaneous emission are taken into account
in the standard way \cite{WallsMilburn}. We compare this dynamics to that of 3 particles in
the 3-site version of (\ref{bosehubbard}) with additional decay terms at rate $\Gamma$
focussing on the expectation value and the fluctuations of the number of photons in one cavity,
i.e. $n_l = \bra a_l^{\dagger} a_l \ket$  and $F_l = \sqrt{\bra (a_l^{\dagger} a_l)^2 \ket - \bra a_l^{\dagger} a_l \ket^2} \: $ ($l = 1,2,3$).

\begin{figure}
\psfrag{data1}{\hspace{0.0cm} \small $U$}
\psfrag{data2}{\hspace{0.0cm} \small $J$}
\psfrag{a}{\hspace{-2.0cm}\raisebox{-0.0cm}{{\bf a}}}
\psfrag{b}{\hspace{-2.0cm}\raisebox{-0.0cm}{{\bf b}}}
\psfrag{c}{\hspace{-2.0cm}\raisebox{-0.0cm}{{\bf c}}}
\psfrag{d}{\hspace{-2.0cm}\raisebox{-0.0cm}{{\bf d}}}
\psfrag{t}{\hspace{-0.4cm}\raisebox{-0.1cm}{\tiny $t$ in $10^{-6}$ s}}
\psfrag{n}{\tiny $n_1$}
\includegraphics[width=8cm]{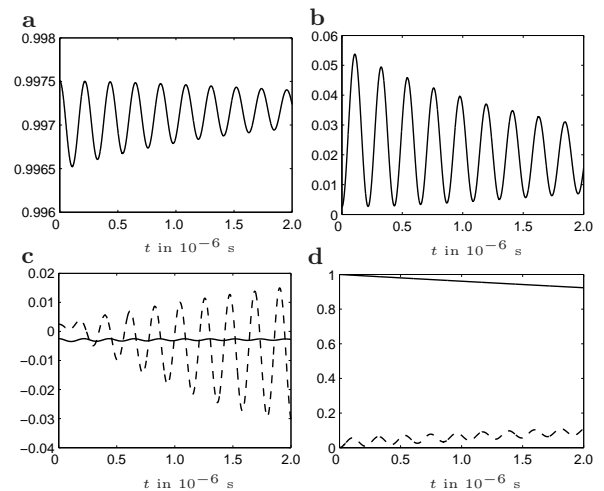}
\caption{\label{Mottplot} Mott insulator state for three polaritons in three cavities:
{\bf a}: photon number $n_1$, {\bf b}: photon number fluctuation $F_1$, {\bf c}: differences between full and effective BH-model, $\delta n_1$ (solid line) and $\delta F_1$ (dashed line) and {\bf d}: $n_1$ and $F_1$ as given by a solution of a master equation for the three site BH-model (\ref{bosehubbard}) including
damping $\Gamma$.}
\end{figure}

Fig. \ref{Mottplot} shows the dynamics of 3 photons in 3 cavities with periodic
boundary conditions. Figures \ref{Mottplot} {\bf a} and {\bf b} show the number of photons, $n_1$, and number fluctuation, $F_1$, in one cavity according to the full model.
Figure \ref{Mottplot} {\bf c} shows the differences between full and effective BH-model,
$\delta n_1 = \left[n_1\right]_{\text{cavities}} - \left[n_1\right]_{\text{BH}}$ (solid line) and
$\delta F_1 = \left[F_1\right]_{\text{cavities}} - \left[F_1\right]_{\text{BH}}$ (dashed line).
All these three plots show results for one single trajectory of
a quantum jump simulation \cite{PK98}. The abatement of the norm showed that the damping rate $\Gamma$
accurately describes the model's losses (deviations less than 5\%). Figure \ref{Mottplot} {\bf d} finally shows $n_1$ and $F_1$ as given by a solution of a master equation for the 3-site BH-model including damping according to $\Gamma$,
which corresponds to the values to be observed in an experiment. There is a probability of $\sim 8$\% to loose the photon in the simulated time range.

\begin{figure}
\psfrag{data1}{\hspace{0.0cm} \small $U$}
\psfrag{data2}{\hspace{0.0cm} \small $J$}
\psfrag{a}{\hspace{-2.0cm}\raisebox{-0.0cm}{{\bf a}}}
\psfrag{b}{\hspace{-2.0cm}\raisebox{-0.0cm}{{\bf b}}}
\psfrag{c}{\hspace{-2.0cm}\raisebox{-0.0cm}{{\bf c}}}
\psfrag{d}{\hspace{-2.0cm}\raisebox{-0.0cm}{{\bf d}}}
\psfrag{t}{\hspace{-0.4cm}\raisebox{-0.1cm}{\tiny $t$ in $10^{-6}$ s}}
\psfrag{n}{\tiny $n_1$}
\includegraphics[width=8.6cm]{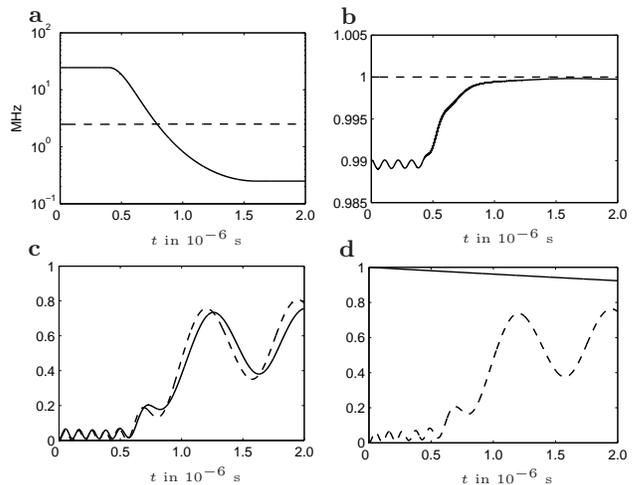}
\caption{\label{transitionplot} Mott insulator to superfluid transition for three polaritons in three cavities:
{\bf a}: On-site potential $U$ (solid line) and hopping $J$ (dashed line), {\bf b}: number of photons in cavity 1, $n_1$ for the full model (solid line) and the effective BH-model (dashed line),
{\bf c}: fluctuations of the photon number in cavity 1, $F_1$ for the full model (solid line) and the effective BH-model (dashed line) and {\bf d}: $n_1$ and $F_1$ as given by a solution of a master equation for the three site BH-model (\ref{bosehubbard}) including damping
$\Gamma$.}
\end{figure}

\paragraph{Mott insulator to superfluid transition:} 
Due to the high Q-factors of present day micro-cavities our approach even allows to observe quantum phase transitions for interacting photons. We consider here the transition from the Mott insulator state to a superfluid regime. The on-site potential $U$ can be decreased by increasing the intensity of the driving laser and hence $\Omega$. Fig. \ref{transitionplot} shows that the local
photon number fluctuations are very small in the Mott phase and increase as $U$ becomes smaller than $J$
and the system is driven into the superfluid phase. The parameters are the same as in figure \ref{Mottplot} except for $\Delta = - 2.5 \times 10^{9} \text{s}^{-1}$ and $J = 2.5 \times 10^{6} \text{s}^{-1}$. $\Omega$ is initially $\Omega_i = 10 \sqrt{N} g_{13}$ and is increased to $\Omega_f = 100 \sqrt{N} g_{13}$ to drive the system through the transition.

Let us stress again, that figures \ref{Mottplot} and \ref{transitionplot} show numbers and number fluctuations for photons and not polaritons.
The photon number statistics in one cavity can be measured via a procedure described in 
\cite{I02} and \cite{HBP06} (c.f. \cite{HMH04} for applications).

\paragraph{Experimental requirements}
To generate a non-linearity that sufficiently exceeds the decay rates such that the resulting Mott state
has a long lifetime, atom-photon couplings should exceed cavity decay rates by one or two orders of magnitude. Promising candidates for an implementation are therefore toroidal micro-cavities
which have a very large Q-factor ($> 10^8$), can be produced in large numbers, positioned with high precision and efficiently coupled to optical fibres as well as to Cs-atoms via their evanescent field \cite{AKS+03}. Photonic crystals represent an appealing longer term
alternative as they offer the possibility for the fabrication of large arrays of cavities in lattices or networks \cite{BHA+05}. Other recently developed micro-cavities, such as \cite{THE+05}, have very high cooperativity factors but are less suited for the present purpose because of their relatively high cavity decay rates.

In contrast to \cite{HBP06} and \cite{ASB06} there is no restriction on $N$ as excitations are photonic and the requirement that level 2 is long lived can be relaxed as it is very weakly occupied. Therefore, circuit cavities interacting with a Cooper pair box \cite{WSB+04} which would have very low cavity decay if operated in a row ($Q \sim 10^6$) are very suitable for an implementation, too. Here, a ratio of $U / \Gamma \sim 15$
could be achieved in current experimental devices. Current measurement schemes in circuit cavities would however require open end ports implying higher cavity decay.

Variations in $g_{13}$ or $N$ for different cavities can be compensated by individually adjusting the laser driving $\Omega$ for each cavity. Relative fluctuations in $g_{24}$ of as much as 50\% are permitable as $U \propto g_{24}$ and the $U$ of each cavity would still be much larger than $\Gamma$ but much lower than $|\mu_{\pm} - \mu_0|$, ($|\mu_{\pm} - \mu_0| \sim 10^{12} \text{s}^{-1}$ for the parameters of figure \ref{Mottplot}). For the same reasons relative fluctuations of 50\% in $\Delta$ are tolerable. A possible detuning $\delta_C$ between neighboring cavities reduces the photon hopping by a factor $J/\delta_C$, thus even favoring the Mott state, and is harmless if the energy separation between polariton species persists, i.e. $\delta_C \ll |\mu_{\pm} - \mu_0|$ ($\delta_C < 10^{10} \text{s}^{-1}$ for micro-toroids).
However only for $|U| > |\delta_C|$, the non-linearity $U$ is the only origin of the immobility of the photons.
Detunings from the two photon resonances with the driving laser generate effective chemical potentials, see above.
Variations in $J$ are tolerable as long as $J \ll |\mu_{\pm} - \mu_0|$.
We note that signatures of the photonic Mott state, such as $F_l \approx 0$ in contrast to the superfluid
regime, can already be observed for as little as two coupled cavities.

The demonstration of the strong coupling regime in micro-toroidal cavities \cite{AKS+03} was done with atoms falling through the interaction region, where the interaction times were about $2\mu$s. For an experiment these cavities could thus be loaded with photons forming a photonic superfluid that are stored long enough due to the high Q. The atoms are then dropped along the cavities. As they approach the evanescent field, their interaction with it increases and the system is driven into the Mott state where it could remain for the duration of the interaction, i.e. $2\mu$s which is enough for our proposal.

\paragraph{Conclusions:}
We have shown that photon non-linearities in EIT systems can be much larger than previously thought.
Our approach does not use additional technology such as lasers etc but demonstrates that certain constraints which were previously thought to be necessary may be relaxed.
With the non-linearities we predict, Mott insulator states and the quantum phase transition
between superfluid and Mott insulator become observable for photons trapped in high-Q micro-cavities.
  
This work is part of the QIP-IRC supported by EPSRC (GR/S82176/0),
the Integrated Project Qubit Applications (QAP) supported by the IST directorate
as Contract Number 015848 and was also supported by the Alexander
von Humboldt Foundation and the Royal Society.

\end{document}